\documentclass[twocolumn]{aastex701}

\begin{document}

\title{Formation and Eruption of Filament Channel in Solar Active Region 12975: Insights from Observations and Simulations of Magnetic Field Evolution  }

\author[0009-0000-3793-0779,sname='Dinesh Mishra']{Dinesh Mishra}
\affiliation{Udaipur Solar Observatory, Physical Research Laboratory, Dewali, Badi Road, Udaipur 313 004, India}

\affiliation{Department of Physics, Indian Institute of Technology Gandhinagar, Palaj, Gandhinagar 382355, India
}
\email[show]{dineshm@prl.res.in}  

\author[0000-0003-4433-8823,sname=P., gname=Vemareddy]{P. Vemareddy}
\affiliation{Indian Institute of Astrophysics, II Block, Koramangala, Bengaluru-560 034, India}
\email{vemareddy@iiap.res.in}

\author[0000-0002-7054-5669,sname='Brajesh Kumar']{Brajesh Kumar}
\affiliation{Udaipur Solar Observatory, Physical Research Laboratory, Dewali, Badi Road, Udaipur 313 004, India}

\email{Brajesh@prl.res.in}
 
\begin{abstract}

We studied the magnetic field evolution of active region (AR) 12975 using a time-dependent magnetofrictional (TMF) model. This AR produced two consecutive CMEs associated with M-class flares on March 28, 2022. The AR exhibited a simple bipolar configuration, with new bipolar flux emerging from March 27. These emerging flux regions evolved through shear motions, forming a filament-channel that ultimately erupted on March 28 at 12:00 UT. The simulation, initialized at 12:00 UT on March 26, is driven by electric fields derived from a time-series of photospheric vector-magnetograms.  It reproduces the observed coronal evolution, including the gradual development of a sigmoidal, twisted flux rope (FR) over approximately 50 hours. The modeled temporal evolution of magnetic energy and helicity within the computational domain is consistent with the observed injection of both quantities. Furthermore, the ratio of current-carrying to total relative helicity reaches 0.23 at the time of observed eruption, however the torus-unstable regime is attained when the helicity ratio reaches 0.32, approximately 7 hr after the observed eruption. Notably, the FR forms adjacent to pre-existing magnetic fields, and a substantial portion of the coronal structure does not belong to the FR system. Consequently, the derived helicity thresholds vary and deviate from the proposed value of 0.29. While reproducing filament formation with high morphological accuracy, this study underscores the quantitative challenges involved in modeling and evaluating the eruptive behavior of different ARs.

\end{abstract}


\section{Introduction} \label{Intro}

Solar flares and coronal mass ejections (CMEs) are the most energetic manifestations of solar activity and major drivers of space weather \citep{kilpua2017bib5, klimchuk2001theory}. The coronal magnetic fields are anchored at the solar surface, where slow underlying motions progressively stretch and twist the fields over time \citep{vanDrielGesztelyi2015}. Through these processes, magnetic energy and helicity are continuously injected into the coronal volume, driving the fields towards increasingly non-potential states (e.g., \citealt{Vemareddy2012, Vemareddy2015}). When a flare or CME occurs, this stored magnetic energy is rapidly converted into heat, radiation, and kinetic energy, producing large-scale eruptions that can significantly impact the heliosphere \citep{Forbes2000, Chen2011}. The energy powering these events is stored in specific magnetic structures, particularly magnetic flux ropes, that are coherent bundles of twisted magnetic field lines carrying substantial electric current and helicity \citep{Cheng2017OriginAS}. Therefore, understanding the gradual evolution of the coronal magnetic field and the formation of flux ropes is essential for explaining the onset and dynamics of solar flares and CMEs \citep{Green2018}.

Because the three-dimensional coronal magnetic fields cannot be routinely measured directly \citep{lin2000bib11}, the extrapolation techniques and numerical modeling play a crucial role in reconstructing coronal magnetic fields and tracking their temporal evolution using photospheric vector magnetograms (e.g., \citealt{Gary1989, Sakurai1989, Schrijver2008}). Among these, nonlinear force-free field (NLFFF) extrapolations incorporate currents and have been widely used to construct three-dimensional coronal magnetic fields from photospheric vector magnetograms (e.g., \citealt{DeRosa_2009, Jiang2014_nlfff, Vemareddy2018_3DMagstr, Vemareddy2021}). However, such approaches generally provide static snapshots and are therefore limited in capturing the continuous, time-dependent buildup of magnetic stress and helicity in the corona. Because the pre-eruptive evolution of ARs occurs over timescales of hours to days and remains largely quasi-static \citep{Vemareddy2014_quasi_stat}, data-driven and time-dependent modeling approaches provide a more physically consistent framework for investigating the gradual evolution of the coronal magnetic fields. It is during this prolonged evolution that magnetic shear and twist accumulate, filament channels form, and flux ropes develop, ultimately determining whether an active region (AR) remains confined or produces an eruption. Within this context, the time-dependent magnetofrictional (TMF) model provides an appropriate and computationally efficient framework for studying the quasi-static evolution of the coronal magnetic field. The magnetofrictional method \citep{Yang1986} is a well-established approach for computing NLFFF configurations by relaxing the initial magnetic fields towards force-free equilibrium. In its time-dependent form \citep{vanBallegooijen2000_tmf}, the TMF approach evolves the coronal magnetic field through a sequence of force-free states while incorporating continuous driving by photospheric electric fields \citep{Cheung2012}.  This enables the model to capture the gradual accumulation of magnetic helicity and free energy, the formation of magnetic flux ropes, and their slow, quasi-static rise prior to eruption \citep{Pomoell2019, Vemareddy_2024}. By adding a non-inductive component--regulated by an ad hoc parameter--to the inductive part of the electric field, an extra degree of non-potentiality can be introduced into the coronal magnetic fields, enabling it to reproduce pre-eruptive structures such as filaments, sigmoids, or EUV hot channels, which are challenging to model with NLFFF extrapolations.  

TMF simulations provide a framework to examine the evolution of coronal magnetic fields and then assess the conditions for eruptive potential of the AR. An important question is whether a helicity threshold can be identified that distinguishes eruptive from stable configurations which was investigated by  \citet{Pariat2017} and \citet{Zuccarello2018}, based on idealized zero-$\beta$ simulations. They had proposed that the ratio of current-carrying helicity to relative helicity ($H_J/H_V$) serves as a  proxy for eruptivity, with a threshold near $\sim$0.29 associated with the onset of the torus instability. Subsequent data-driven TMF simulations have applied this idea across multiple ARs, each offering case-specific insights into \textbf{magnetic flux rope (FR)} formation and subsequent eruption.  \citet{Pomoell2019} simulated AR~11504, which produced an M1.9 flare and halo CME on 2012 June 14, and found that only the simulation with the highest relative helicity content resulted in an eruptive flux rope, highlighting helicity as a key discriminator. The study by \citet{Price2019} modeled the highly flare-productive AR~12673, capturing the X9.3 flare on 2017 September 6, and reported that $H_J/H_V$ approached $\sim$0.17 but was never exceeded, suggesting a possible upper bound for stable configurations. \citet{DjPrice_2020} extended this analysis to AR~12473, which produced an M1.9 flare and CME on 2015 December 28, and identified the torus instability as the eruption mechanism when the flux rope entered a region where the decay index exceeded $n > 1.5$. In an isolated emerging AR, \citet{Lumme2022} simulated AR~11726 and found that a negatively twisted flux rope formed despite the region’s dominant positive helicity, with the $H_J/H_V$ ratio exhibiting peaks during flux rope formation. More recently, \citet{Vemareddy_2024} investigated the AR~11429, magnetic evolution capturing the formation of sigmoid FR exhibiting the eruptive signatures, and demonstrated the capability of TMF to reproduce the pre-eruptive structures in unprecedented detail. Similarly, \citet{Vemareddy2026} analyzed another AR~13500, which produced an M9.8 flare and halo CME on 2023 November 28, and found that $H_J/H_V$ increased from $\sim$0.13 during FR formation to $\sim$0.30 at eruption. At this stage, the FR core resided in a region where the decay index exceeded $n > 1.5$, indicating the onset of torus instability.  Collectively, these studies suggest that the helicity ratio threshold identified in idealized models may also be applicable to data-driven simulations of real ARs. However, the extent to which $H_J/H_V \gtrsim 0.3$ represents a universal eruptivity criterion remains an open question, particularly given the diversity of ARs properties and evolutionary pathways. This motivates further investigation using TMF simulations across a broader range of solar ARs.

In this study, we apply a data-driven TMF model to investigate the magnetic evolution of active region 12975, which produced two CMEs within an eight-hour interval on March 28, 2022. We focus on the evolution of magnetic fields during the first eruption associated with filament eruption.  The objectives of this work are: (1) to analyse the observed magnetic evolution in conjunction with the multi-wavelength observations of the AR, (2) to simulate the coronal magnetic evolution, capturing the filament formation and its eruption and (3) to assess the simulated magnetic evolution for its eruptive potential measures.  We organize the paper as follows: Section 2 presents an overview of the event, spacecraft geometry, and the observational data. Section 3 describes the evolution of the photospheric magnetic fields and their associated coronal evolution, along with the results of the data-driven TMF simulations, including the formation and rise of the magnetic flux rope, and the evolution of magnetic energy and relative helicity in the corona. Section 4 summarizes the main conclusions of this study.

\section{Observations and Overview of AR 12975} \label{Sec:MultiObs}

The NOAA AR 12975 appeared at the eastern solar limb (N13E51) on March 24, 2022, and exhibited significant eruptive activity throughout its disk passage. This study focuses on the period from March 26 to 29, when the region was located near the central meridian, providing a favorable vantage point for Earth-observed instruments and minimizing projection effects in the acquired images of this active region.  On March 28, 2022, this region produced two successive CMEs around 11:25 UT and 17:32 UT within an eight-hour window, each associated with an M-class flare, making the AR an ideal target for studying pre-eruptive magnetic evolution.  The first of these eruptions, the focus of this study, leads to a CME that displayed a classic three-part structure in the low corona: a bright core, a dark cavity, and a bright leading edge. 

\begin{figure*}[ht!]
\centering
\includegraphics[width=0.6\textwidth]{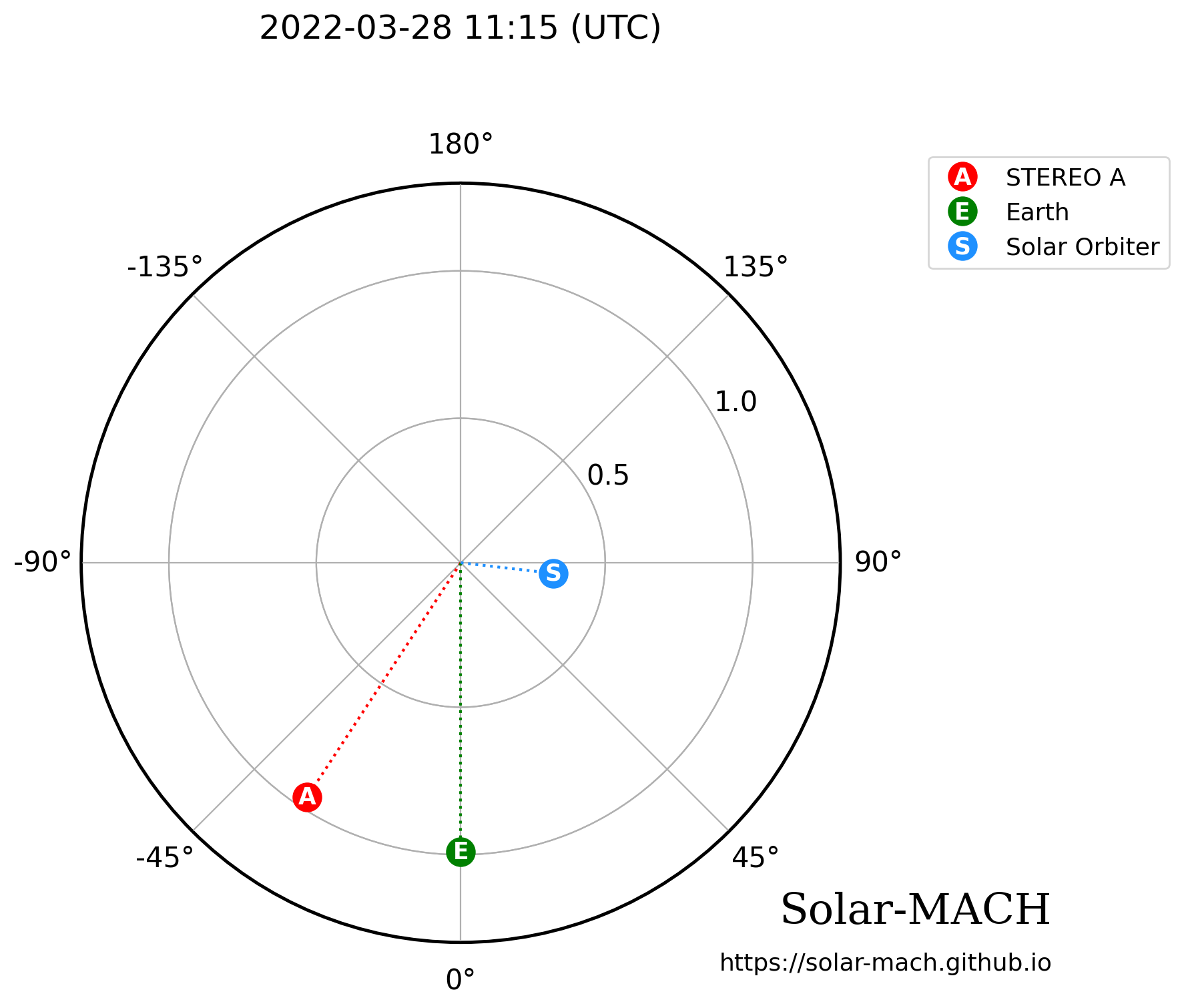}
\caption{Relative positions of SDO, Solar Orbiter (SolO), and STEREO-A during the March 28, 2022 filament eruption. This geometric configuration enabled multi-perspective observations from three distinct vantage points. This visualisation is created using Solar MACH \citep{gieseler2023bib12}.}
\label{fig:spacecraft_positions}
\end{figure*}

This eruption was observed simultaneously by multiple solar space missions located in different perspective views, as shown in Figure~\ref{fig:spacecraft_positions}. The Solar Dynamics Observatory (SDO; \citet{Pesnell2012}) observed the event near disk center from 1~AU. The Solar Orbiter (SolO; \citet{Muller2020}) was located at $\sim$83.5$^{\circ}$ west of Sun-Earth line at 0.33~AU providing an eastern-limb perspective, and The Solar Terrestrial Relations Observatory Ahead (STEREO-A; \citealt{Kaiser2008}) was positioned at $\sim 33^\circ$ east of Earth at 0.97 AU, offering a northwestern on-disk view. Observations from the Sun Earth Connection Coronal and Heliospheric Investigation (SECCHI; \citealt{Howard2008}) instrument onboard STEREO-A were used in this study.

Our analysis primarily utilizes data from Solar Dynamics Observatory (SDO), which provides continuous full-disk solar observations with high spatial and temporal resolution through its two primary instruments: the Helioseismic and Magnetic Imager (HMI; \citealt{Schou2012, Scherrer2012})  and the Atmospheric Imaging Assembly (AIA; \citealt{Lemen2012}). The HMI provides full-disk photospheric observations, including Dopplergrams, continuum intensity maps, and both line-of-sight and vector magnetic field measurements at a resolution of $1''$. In this work, we utilize the \texttt{hmi.sharp\_cea\_720s} data product \citep{Bobra2014}, which provides disambiguated vector magnetic field observations in heliographic coordinates remapped to disk center using the cylindrical equal-area (CEA) projection, with a cadence of 12 minutes and plate scale of $0.5''$ per pixel. The Atmospheric Imaging Assembly (AIA; \citealt{Lemen2012}) captures full-disk images in multiple extreme ultraviolet (EUV) and ultraviolet (UV) channels, sampling plasma temperatures from the upper chromosphere to the hot corona. From \citet{Lemen2012}, the EUV channels are sensitive to plasma temperatures ranging from $\sim 6 \times 10^{4}\,\mathrm{K}$ to $\sim 2 \times 10^{7}\,\mathrm{K}$, while the UV channels extend the coverage to lower temperatures. The instrument provides images with a pixel scale of $0.6^{\prime\prime}$ per pixel and a cadence of 12\,s, enabling detailed tracking of coronal dynamics.

\begin{figure*}[htbp]
    \centering
    \includegraphics[width=\textwidth]{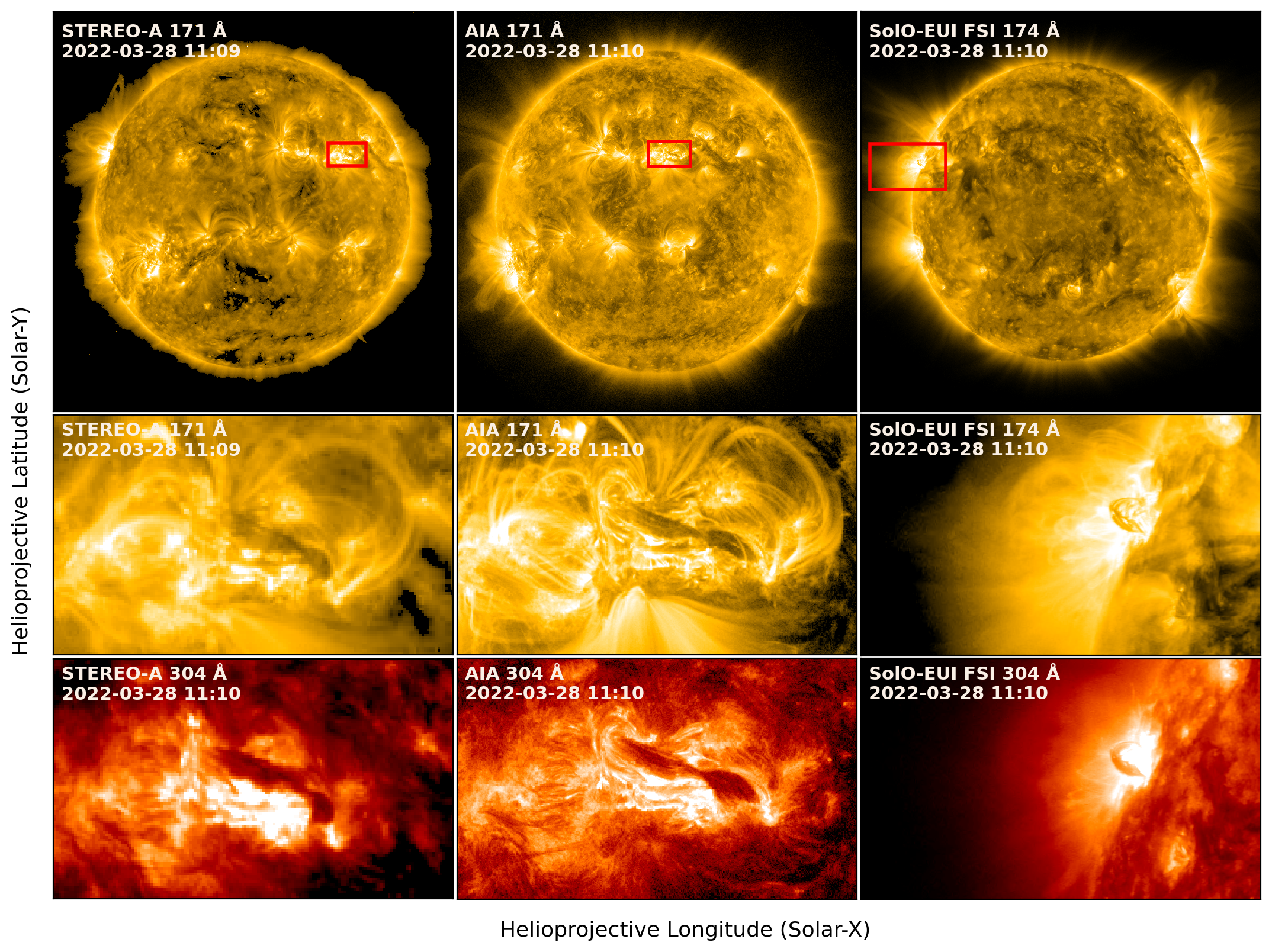} 
    \caption{Multi-viewpoint EUV observations of the AR 12975 on March 28, 2022 from STEREO-A/EUVI, SDO/AIA, and Solar Orbiter/EUI-FSI. The top row shows full-disk images in the 171~\AA \ (STEREO-A and AIA) and 174~\AA\ (SolO-EUI FSI) channels, where the red rectangles indicate the region of interest. The middle row presents zoomed-in views of the same region in the corresponding coronal wavelengths, highlighting the filament channel and surrounding loop structures. The bottom row displays zoomed-in images in the 304~\AA\ channel, showing the cooler filament material. The three distinct vantage points provide complementary information on the geometry and three-dimensional structure of the filament.}
    \label{fig:multiinstrument}
\end{figure*}

The pre-eruptive feature is the filament, which is well captured in near-simultaneous EUV observations from STEREO-A/EUVI, SDO/AIA, and Solar Orbiter/EUI-FSI, which together provide three widely separated viewing angles of the same structure as illustrated in Figure~\ref{fig:multiinstrument}. The top row shows full-disk images in the 171~\AA\ and 174~\AA\ passbands, with the region of interest indicated by red boxes. From the Earth-based AIA perspective, the filament is located near the disk center and appears as an extended dark channel, while EUI/SolO observes the same structure near the east limb, offering a partially side-on view. STEREO-A captures the filament close to the west limb, providing a complementary perspective. The middle row of Figure~\ref{fig:multiinstrument} presents zoomed-in views in the same coronal passbands, revealing the detailed morphology of the filament and the overlying arcade. The loops appear relatively compact and low-lying in the AIA view, whereas the SolO perspective shows a more vertically extended configuration, consistent with an elevated filament axis. The bottom row of Figure~\ref{fig:multiinstrument} displays the corresponding 304~\AA \ images, which trace the cooler filament plasma and clearly outline its structure.

\begin{figure*}[htbp]
    \centering
    \includegraphics[width=\textwidth]{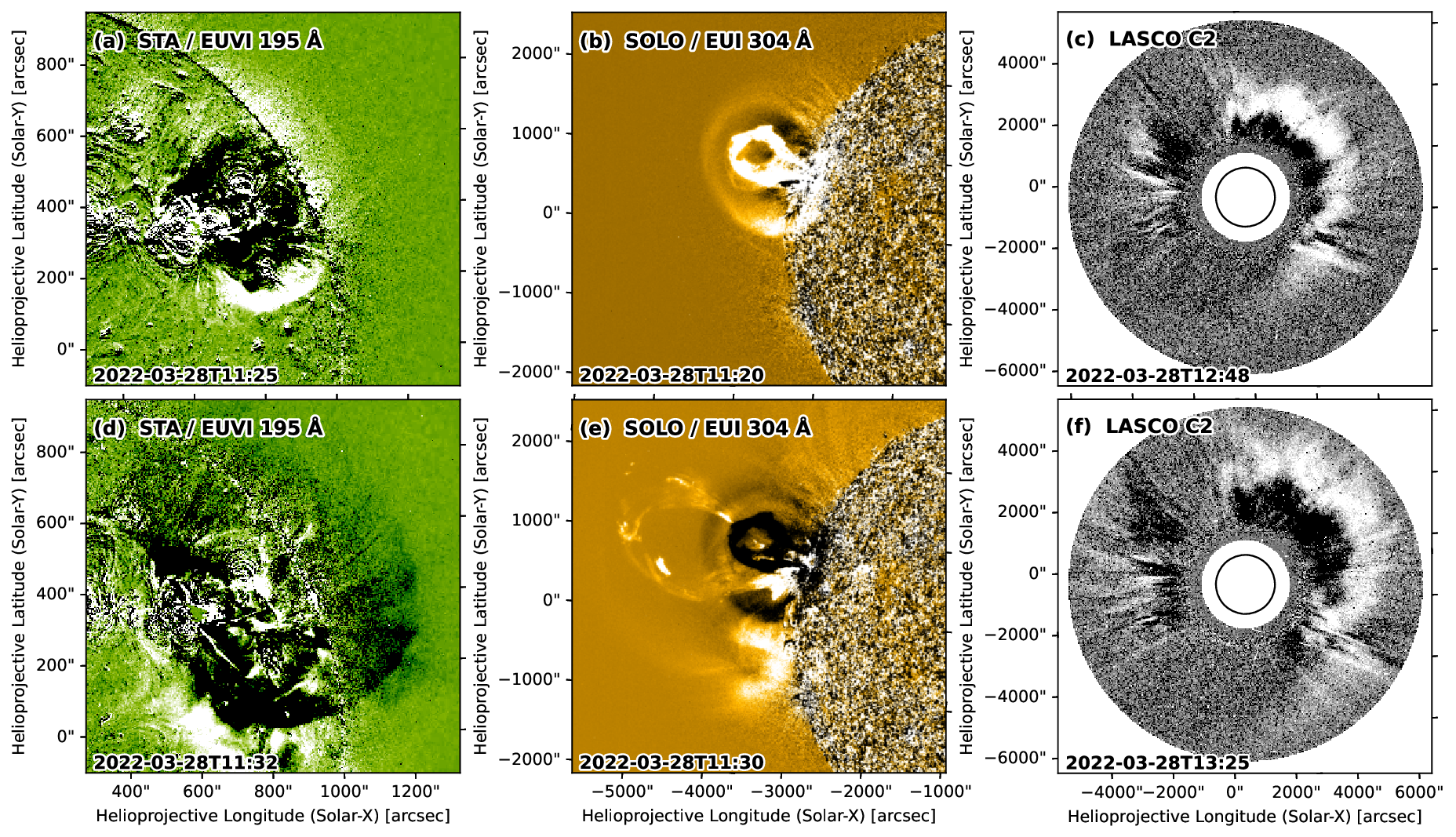} 
    \caption{Multi-view observations of the eruption on March 28, 2022 from three heliospheric vantage points, showing the associated CME propagating into the outer corona. Panels (a) and (d) show STEREO-A/EUVI 195~\AA\  running-difference images, highlighting the expansion of CME the early eruption phase. Panels (b) and (e) present SOLO/EUI-FSI 304~\AA\  running-difference images, where the erupting filament core and surrounding expanding front are clearly visible. Panels (c) and (f) display SOHO/LASCO C2 coronagraph images at later times, showing the outward-propagating CME with a distinct three-part structure.}
    \label{fig:CME}
\end{figure*}

Following the filament eruption, the associated CME propagated into the outer corona and interplanetary space. Running-difference images from STEREO-A/SECCHI EUVI 195~\AA\ (Figure~\ref{fig:CME}a--d) track the early expansion of the CME front. In the Solar Orbiter 304~\AA\ running–difference images (Figure~\ref{fig:CME} b and e), the erupting filament appears as a dark, rising core surrounded by an expanding bright front. The CME exhibits a well-defined three-part structure, consisting of a bright leading front of compressed plasma, a central cavity, and a bright filament or prominence core. This morphology indicates the outward expansion of plasma and the formation of the CME cavity during the eruption. 

 As the CME propagates outward to larger heliocentric distances, it is observed in white light by the LASCO coronagraph onboard \textit{SOHO} \citep{1995SoPh..162..357B}, as shown in Figure~\ref{fig:CME}(c-f). According to the LASCO CME catalogue \citep{2004JGRA..109.7105Y}, the event is classified as a halo CME, with a linear speed of $\sim700~\mathrm{km~s^{-1}}$ and a reported acceleration of $\sim -47~\mathrm{m~s^{-2}}$. 

\section{Analysis and Results}
\label{sec:Res}

\subsection{Observed magnetic field evolution}

Using HMI vector magnetic field observations (\texttt{hmi.sharp\_cea\_720s}) \citep{Bobra2014}, the magnetic field evolution of AR\,12975 was studied from 12:00~UT on March 26  until 12:00~UT on March 29. 
Throughout the observing period, AR\,12975 maintained a relatively simple magnetic configuration, classified as a $\beta$-type region with a D-type sunspot group, according to the Kanzelhöhe Observatory (KSO; \cite{Pötzi_2021}). The region exhibited a developing bipolar structure without dominant sunspots, characterized by scattered, small-scale magnetic elements.

\begin{figure*}[htbp]
    \centering
    \includegraphics[width=0.9\textwidth]{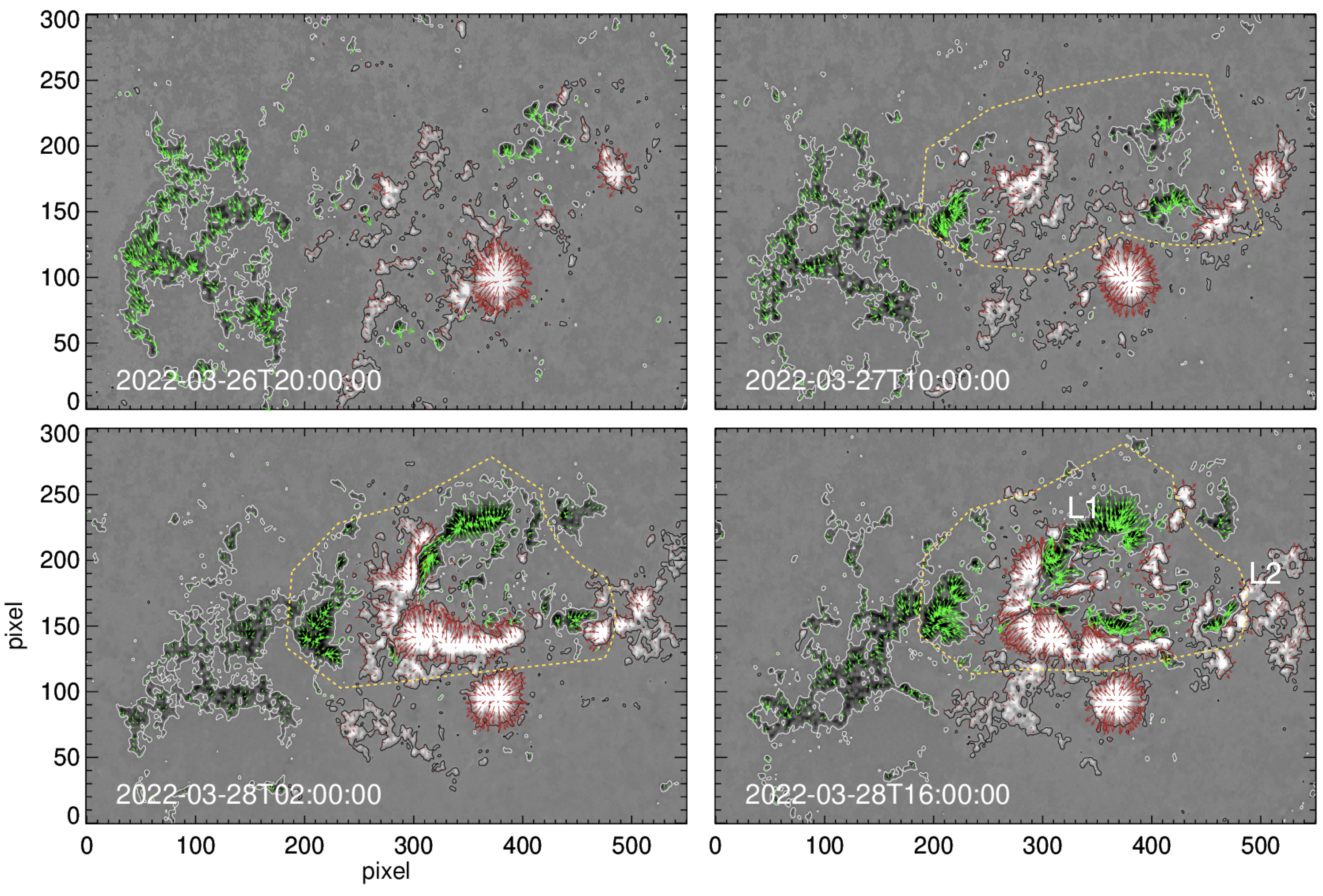}
    \caption{Evolution of the vector magnetic fields in AR\,12975 at four representative times during March 26 to 28, 2022. The background shows the vertical magnetic field components (positive polarity in white, negative in black). Overplotted arrows represent the transverse magnetic fields, colored by vertical polarity (red in positive, green in negative regions). The yellow polygon encloses the magnetic fluxes being emerged. Flux regions L1 and L2 denote the opposite footpoints of the filament.  }
    \label{fig:vectorB}
\end{figure*}

Figure~\ref{fig:vectorB} illustrates the photospheric vector magnetic field evolution at four different times. At 20:00 UT on March 26, the AR comprises a leading positive polarity with a dominant sunspot and fragmented following polarity regions. Then, bipolar flux concentrations start emerging, which become evident in the central region, emerging by 10:00 UT on March 27. They further grow in size over time, as seen on March 28. This newly emerged region (enclosed by a yellow polygon) consisted of interacting opposite polarities, with a sheared polarity inversion line; however, its stretch is small compared to other ARs (eg, NOAA 11158, 11429, 12371, 12673,  \citep{Vemareddy2019}). As the evolution progresses, the central region develops strong horizontal magnetic fields by 16:00~UT on March 28, exhibiting clear signatures of magnetic shear that influences the coronal configuration, leading to the formation of a filament structure approximately one day before the eruption. During the early stages, coronal loops observed in AIA~171~\AA\ were weakly sheared, while intermittent brightenings in AIA~304~\AA\ indicated ongoing flux emergence. With continued flux emergence and their shear motion, the coronal fields became progressively more sheared, which probably leads to the formation of a filament channel as discernible in AIA~304~\AA\ around 03:30~UT on March~28. The footpoints of the filament legs are marked L1 and L2 in the last panel of Figure~\ref{fig:vectorB}. 

The filament system remained stable until a confined C-class flare at about 09:55~UT triggered a reconfiguration of the coronal magnetic fields, during which the filament elongated and its morphology changed. Shortly thereafter, the filament accelerated rapidly around 11:20~UT and erupted fully by approximately 11:25~UT, producing a fast halo CME and M4.0 class flare.

\begin{figure*}[ht]
    \centering
    \includegraphics[width=0.7\textwidth]{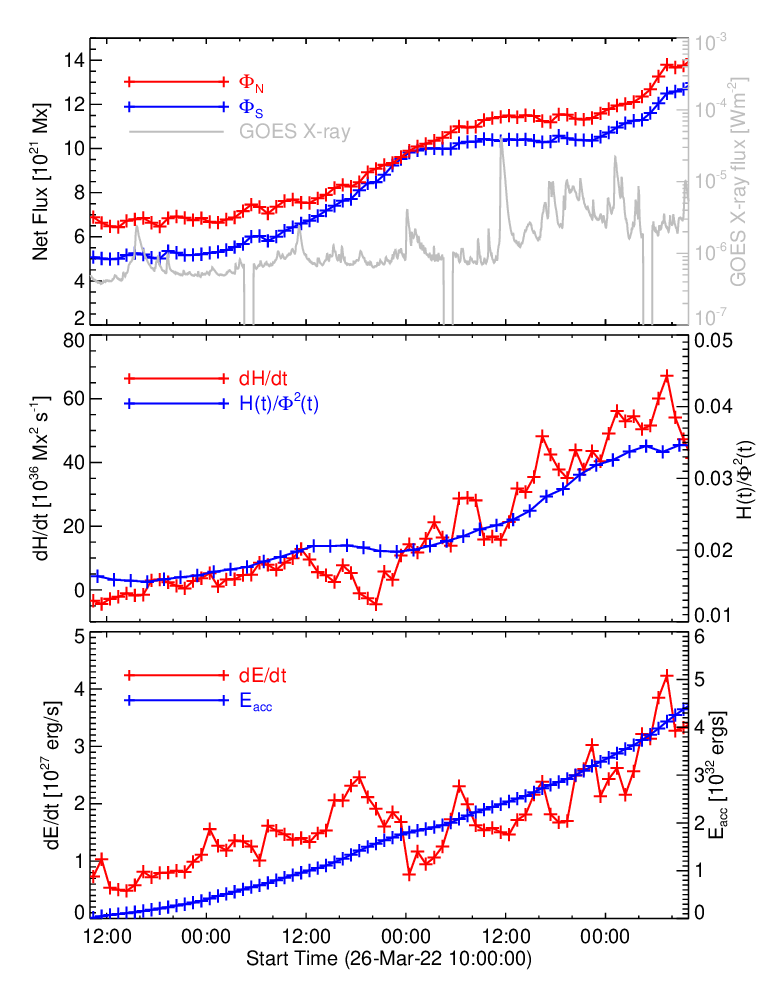}
    \caption{{\bf Top:} Temporal evolution of the magnetic flux in AR~12975, with positive (red) and negative (blue) polarities. 
    The GOES soft X-ray light curve (grey) is also overplotted for comparison. 
    {\bf Middle:} Temporal profile of the magnetic helicity injection rate (red) and the accumulated helicity (blue) in AR~12975. 
    {\bf Bottom:} Temporal evolution of the magnetic energy injection rate (red) and the accumulated magnetic energy (blue) in the corona.}
    \label{fig:fluxes}
\end{figure*}

The evolution of the photospheric magnetic flux in AR 12975 was analysed from 10:00 UT on March 26 to 10:00 UT on March 29. The magnetic flux $\Phi = \sum B_{z} \cdot dA$ was calculated by integrating the vertical magnetic field component $B_z$ over the AR area. The top panel of Figure ~\ref{fig:fluxes} shows the time evolution of the positive ($\Phi_N$, red) and negative ($\Phi_S$, blue) magnetic fluxes, overplotted with the disk-integrated GOES \citep{Woods2024} soft X-ray flux  (grey curve) for the temporal evolution of the flare. Both polarities exhibit a steady, nearly simultaneous increase throughout the observation period. The positive flux rises from approximately $7 \times 10^{21}$ Mx at 10:00 UT on March 26, 2022  to $1.2 \times 10^{22}$ Mx by 10:00 UT on March 29, 2022 while the negative flux increases from about $5 \times 10^{21}$ Mx to $1.3 \times 10^{22}$ Mx over the same time interval. This co-temporal growth indicates continuous magnetic flux emergence beginning on March 26, consistent with the appearance, growth, and interaction of bipolar magnetic regions. A modest flux imbalance is present, with the positive flux remaining slightly higher than the negative flux during most of the period. Continuous flux emergence results in the accumulation of magnetic energy and helicity, which drives the formation of a non-potential field (filament channel) that subsequently produces eruptions in this AR. The GOES soft X-ray flux (grey curve, right axis in top panel of Figure~\ref{fig:fluxes}) shows seven C-class and four M-class flares on March 28. The two prominent peaks in the X-ray flux correspond to the two M-class flares around 11:25~UT and 19:00~UT, which are associated with the CMEs that are the focus of this study.

To understand how the emerged magnetic flux was converted into eruptive energy, we analyse the injection of magnetic helicity and energy into the corona. As the AR emerges, horizontal plasma motions at the photosphere shuffle the footpoints of magnetic flux tubes. Under flux-freezing conditions, this shuffling and braiding of the footpoints increases the shear and twist of the coronal magnetic field \citep{priest2002bib10,DemoulinBerger2003, Demoulin2007}.  The injection rate of relative magnetic helicity across a surface~$S$ is given by \cite{Berger1984}:
\begin{equation}
\frac{dH}{dt}
= 2 \int_{S}
\left[
(\mathbf{A}_{\mathrm{p}} \cdot \mathbf{B}_{t})\, v_{n}
-
(\mathbf{A}_{\mathrm{p}} \cdot \mathbf{v}_{t})\, B_{n}
\right]
\, dS,
\label{eq:helicity}
\end{equation}
where $\mathbf{A}_{\mathrm{p}}$ is the vector potential of the potential magnetic field, and the subscripts t, n refer to transverse, vertical components of magnetic ($\mathbf{B}$) and velocity ($\mathbf{V}$) fields.  The first term represents helicity injection ($dH/dt$) due to the emergence of twisted flux, while the second term quantifies injection due to shearing motions. The photospheric velocity field was derived using the Differential Affine Velocity Estimator for Vector Magnetograms (DAVE4VM, \citealt{Schuck2008}), applied to a time series of HMI vector magnetograms. 

\begin{figure*}[ht]
    \centering
    \includegraphics[width=0.7\textwidth]{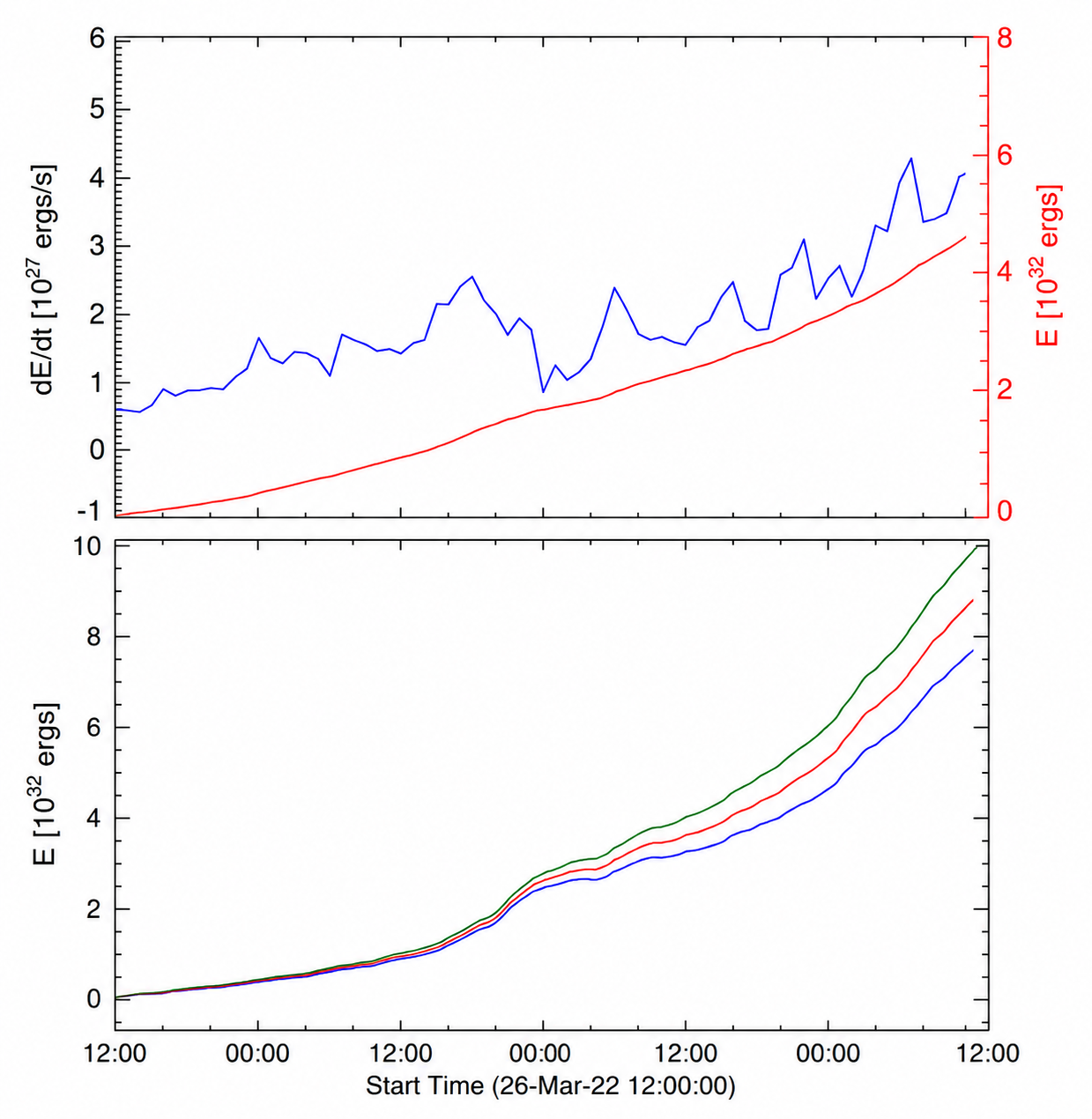}
    \caption{Top: Temporal evolution of the magnetic energy injection rate (blue) and the accumulated magnetic energy (red) observationally. 
    Bottom: Temporal evolution of magnetic energy for \(U = 120~\mathrm{m\,s^{-1}}\) (blue), \(U = 150~\mathrm{m\,s^{-1}}\)  (red), \(U = 180~\mathrm{m\,s^{-1}}\) (green)  .}
    \label{fig:enegy_injection}
\end{figure*}

The computed $dH/dt$ (Figure~\ref{fig:enegy_injection}(b)) shows an increasing trend throughout the observation window. By the eruption time at 12:00~UT on March 28, 2022 $dH/dt$ reaches values of $\approx 16 \times 10^{36}$~Mx$^{2}$~s$^{-1}$, and continues to increase thereafter, attaining peak values of $\sim 60 \times 10^{36}$~Mx$^{2}$~s$^{-1}$ toward the end of the observing period on March 29, 2022. The flux motions inject positive helicity into the system at an average rate of $\sim 2 \times 10^{36}$~Mx$^{2}$~s$^{-1}$, resulting in a total accumulated helicity of $\approx 2.3 \times 10^{42}$~Mx$^{2}$ prior to the eruption.

Similarly, the magnetic energy injection rate is quantified by the Poynting flux \citep{Kusano2002}:

\begin{equation}
\frac{dE}{dt}
= \frac{1}{4\pi} \int_{S}
\left[
(\mathbf{B}_{t} \cdot \mathbf{B}_{t})\, v_{n}
-
(\mathbf{B}_{t} \cdot \mathbf{v}_{t})\, B_{n}
\right]
\, dS.
\label{eq:poynting}
\end{equation}
By integrating this energy flux over time, we calculated the accumulated magnetic energy in the AR. The average energy injection rate was approximately $1.6 \times 10^{27}$~erg\,s$^{-1}$, with a total accumulated energy of about $~2.5 \times 10^{32}$~erg prior to the eruption.

\subsection{Modeling the magnetic field evolution}
To understand how the observed photospheric energy and helicity injection translate into the formation and eruption of a coronal FR, we employ a data-driven MF simulation. The method was originally proposed by \citet{Yang1986} and has since been successfully applied in numerous studies \citep{Cheung2012, Lumme2017, Pomoell2019}. In this work, we follow the implementation described in \citet{Vemareddy_2024}, which evolves the coronal magnetic fields from an initial potential state using photospheric electric fields as the bottom-boundary driver. 

The simulation is initialized at 12:00~UT on March 26 with a potential magnetic field extrapolated from an HMI vector magnetogram. The coronal volume is discretized on a uniform Cartesian grid of $224 \times 224 \times 120$ points, spanning a physical domain of $245 \times 245 \times 131$~Mm$^{3}$. The temporal evolution is driven at the lower boundary by the photospheric electric fields, derived from a time series of (\texttt{hmi.sharp\_cea\_720s} data product) vector magnetograms and pre-processed following \citet{Vemareddy_2024}. The electric field includes a non-inductive component parameterized by an emergence speed $U$, which controls the injection of magnetic twist, consistent with earlier implementations of this approach.

 The simulation spans 80 hours of coronal evolution, beginning at 12:00 UT on March 26. The photospheric driving—implemented through the electric fields boundary condition—is applied for the first 60 hours and is switched off at 20:00 UT on March 28. For the remaining 20~hours, no further driving is applied, allowing the coronal magnetic fields to evolve from the state reached at the end of the driven phase. This two-stage setup enables the model to capture both the gradual, externally forced buildup of magnetic energy and the subsequent internal dynamical evolution that can naturally lead to an eruption.

We performed a series of simulations by evolving the initial potential field with the time-dependent electric field derived from the ad hoc assumption in Equation~(8) of \citet{Vemareddy_2024}, varying the velocity \(U\) from \(120~\mathrm{m\,s^{-1}}\) to \(180~\mathrm{m\,s^{-1}}\). The Figure~\ref{fig:enegy_injection} (top panel) shows the energy injection rate and the corresponding accumulated energy (red curve) from the DAVE4VM analysis. The bottom panel compares the accumulated magnetic energy over time for three U-values. The electric fields with \(U = 120~\mathrm{m\,s^{-1}}\) yields an energy of \(\sim 7.75 \times 10^{32}\)~erg, within \(\sim 39\%\) of the observed total of \(\sim 4.75 \times 10^{32}\)~erg. The simulations with \(U = 150~\mathrm{m\,s^{-1}}\) and \(U = 180~\mathrm{m\,s^{-1}}\) produce proportionally higher final energies of \(\sim 9 \times 10^{32}\)~erg and \(\sim 1.0 \times 10^{33}\)~erg, respectively. Despite the closer energy match for \(U = 120~\mathrm{m\,s^{-1}}\), a comparison with AIA observations (Section~3.2) revealed that the simulations with higher \(U\) values, particularly \(U = 180~\mathrm{m\,s^{-1}}\), more accurately reproduced the timing and morphological evolution of the observed filament formation and its pre-eruptive rise. In order to reproduce the observed twisted structures, a higher amount of energy injections is needed. Following the reasoning in \citet{Vemareddy_2024}, where a similar factor of \(\sim 2\) difference in injected energy was considered acceptable given methodological uncertainties, we adopted \(U = 180~\mathrm{m\,s^{-1}}\) for our analysis to reproduce the observed structural evolution.

Figure~\ref{fig:fieldlines1} shows a comparison between the simulated magnetic field lines and the corresponding SDO/AIA observations at 24, 30, and 48 hours after the simulation begins at 12:00 UT on March 26, 2022. Initially, the coronal magnetic fields are close to a potential configuration, consistent with EUV images from AIA 171\AA, and reflects the combined structure of preexisting polarities and the emerging fields. As flux emergence proceeds, the coronal field gradually evolve into a more twisted configuration. At 24 hours, the low-lying coronal fields become strongly sheared and align with the polarity inversion line (PIL), while the overlying fields remain weakly twisted and nearly potential. By 30 hours, a twisted core structure forms above the PIL of the emerging fields in the northwestern part of the active region, with the overlying arcade beginning to wrap around it. By 48 hours, a coherent magnetic flux rope (FR) is fully developed, consisting of strongly twisted helical field lines enclosed within less twisted, S-shaped arcades. This results in a sigmoidal magnetic configuration that closely resembles filament structures observed in EUV images.

After sigmoidal FR (filament), the simulated fields evolve further quasi-statistically. The Figure~\ref{fig:fieldlines2} illustrates the three-dimensional evolution of the magnetic structure at 50 hr, 60 hr, and 70 hr illustrating the slow rise motion of the formed FR, which reflects the onset of its eruption. The top row shows the top-view morphology, where the flux-rope system exhibits a progressively enhanced sigmoidal configuration with time. At 50 hours, the FR is well-formed but remains relatively compact. By 60 hours, the FR expands laterally and becomes more coherent, while at 70 hr the sigmoidal structure is further stretched and elongated, indicating continued growth of the system. The middle row shows the corresponding perspective (side) views, which clearly reveal the gradual upward rise of the FR with time. The vertical extent of the structure increases significantly from 50 hr to 70 hr, indicating the quasi-static expansion of the coronal magnetic fields. The field line twist ($T_w=\int_L \frac{\mu_0 J_{\parallel}}{4\pi B} \, dl$) evaluated in the FR cross-section plane is shown in the bottom panels of Figure~\ref{fig:fieldlines2}. Red and blue colors represent negative and positive twist, respectively, scaled to $\pm$2 turns. The FR enveloped by a less sheared field is having positive helicity. The FR core (marked by blue arrows) rises progressively in height over time: at 50 hours, the apex of the central twisted core is approximately 16 Mm, increasing to about 32 Mm at 60 hours and reaching nearly 44.7 Mm at 70 hours. This systematic upward displacement of the twisted core demonstrates onset of eruption which is slow due to inherent quasi-static nature of MF simulations. We note that the height derived from the twist maps corresponds to the location of the twisted core on the diagnostic cut plane and is therefore lower than the FR apex height tracked from the outermost field lines.

During this interval, Solar Orbiter was positioned at a heliolongitudinal separation of approximately $83.3^\circ$ from the Sun--Earth line, providing a near-limb perspective of the erupting filament. In the simulation, the FR forms along the PIL and is anchored in the positive leading polarity of the AR  (c.f., Figure~\ref{fig:vectorB}). To facilitate a direct comparison with the SolO viewpoint, we rendered representative magnetic field lines of the modeled filament, as shown in Figure~\ref{fig:filament_height}. The resulting morphology exhibits a remarkable resemblance to the observed filament structure. From SolO, the projected filament height increases from approximately 21~Mm at 10:10~UT to 35~Mm at 11:10~UT, whereas in the simulation, the apex of the FR rises from about 22.3~Mm at $t=53$~hr to 34.4~Mm at $t=67$~hr. The close correspondence between the modeled and observed heights suggests that the simulation successfully reproduces the large-scale configuration and gradual rise of the erupting structure. However, this comparison should be interpreted with caution, as it is not temporally consistent. The observed filament undergoes a rapid and highly dynamic ascent, whereas the TMF model describes the quasi-static evolution of the coronal magnetic fields. Consequently, the simulation captures the gradual build-up and large-scale morphology of the erupting structure but does not represent the fully dynamic eruption phase \citep{Vemareddy2026}.

\subsection{Assessing the coronal non-potential parameters}
We have also analysed the total relative helicity and energy for the three different runs of $U=120~\mathrm{m~s^{-1}}$, $U=150~\mathrm{m~s^{-1}}$, and $U=180~\mathrm{m~s^{-1}}$ in the computational volume. The Figure~\ref{fig:energy_metric}  shows the temporal evolution of the total magnetic energy ($E_T$), free energy ($E_f$) contained in volume and ratio of free energy to total magnetic energy.  The $E_T$ increases slowly during the first $\sim$30–35 hours for all three simulations, followed by a more pronounced rise thereafter. This early-phase evolution is also consistent with the photospheric magnetic-flux evolution (Figure~\ref{fig:fluxes}). 

By the time of the eruption, the $E_T$ reaches $\sim 4.5 \times 10^{32}$~erg for the case with $U = 180~\mathrm{m\,s^{-1}}$, with the lower-$U$ simulations producing comparable values. These energy estimates are sufficient to power M-class eruptions and fall within the range of coronal magnetic energies inferred for eruptive ARs \citep{Cheung2012, Inoue_2014, Pomoell2019}. The free magnetic energy ($E_f$) shows a similar trend, starting to increase after $\sim$30 hours and reaching $\sim 1.5 \times 10^{32}$~erg at the time of the eruption for the $U = 180~\mathrm{m\,s^{-1}}$ run. This corresponds to a free-energy fraction of about 35\% of the total magnetic energy, which is likely sufficient to initiate the rise of the modeled FR. This value lies within the range reported in previous studies. While some MHD simulations suggest that eruptions can be triggered with free-energy fractions as low as $\sim$5\% \citep{AmariEtAl2003}, TMF models reported higher values, typically in the range 14\%–50\% \citep{Pomoell2019, Price2019, DjPrice_2020, Vemareddy_2024, Vemareddy2026}. Our result of $\sim$35\% is therefore consistent with those reported for TMF simulations. Throughout the 72 hr simulation, the free energy continues to increase even after the eruption time, consistent with the steady rise in total magnetic energy. This behaviour can be attributed to the continued emergence of magnetic flux at lower boundary condition, which continues to inject magnetic energy and helicity into the system for several hours after the observed eruption. 

The ratio of free to total magnetic energy, however, begins to decrease after $\sim 60$ hr  ( bottom panel of Figure~\ref{fig:energy_metric}). This decline occurs because the photospheric driving is switched off at 60 hours, after which the coronal field evolves following further FR slow rise. In the absence of further energy injection, the imposed resistivity in the simulation domain dissipates electric currents, leading to the observed relaxation and reduction in the free-energy fraction.

We have also evaluated the decomposed components  of the relative helicity ($H_R = H_j + H{pj}$, \citet{Berger2003}), the current-carrying (non-potential) component, $H_j = \int_V \left( \mathbf{A} - \mathbf{A}_p \right) \cdot \left( \mathbf{B} - \mathbf{B}_p \right), dV$  $H_j$, and mutual component, $H_{pj} = 2 \int_V \mathbf{A}_p \cdot \left( \mathbf{B} - \mathbf{B}_p \right), dV $. The term $H_j$ represents the helicity associated with electric currents and the non-potential part of the field which is a crucial parameter to quantify the magnetic structure that has eruptive potential, while $H{pj}$ reflects the coupling between the potential and non-potential components. The temporal evolution of these components are shown in Figure~\ref{fig:helicity_ratio}. Both the current-carrying helicity $H_{j}$ (top panel) and the total relative magnetic helicity $H_{v}$ (middle panel) remain relatively low during the initial $\sim 35$ hr of the simulation, after which they exhibit a steady increase throughout the rest of the simulation. This behaviour reflects the formation and buildup of FR in the coronal volume driven by the imposed electric field conditions. The shaded region indicates the time interval corresponding to the most probable eruption window in the simulation. At this stage, the current-carrying helicity reaches $~ 1.5 \times 10^{42}~\mathrm{Mx^{2}}$ (for  $U = 180~\mathrm{m\,s^{-1}}$) while the total relative magnetic helicity attains a value of  $~ 4.2 \times 10^{42}~\mathrm{Mx^{2}}$.
The ratio of current-carrying helicity to total relative helicity, $\left|H_j\right| / \left|H_V\right|$, is presented in the bottom panel of Figure~\ref{fig:helicity_ratio}. In line with the evolution of $H_j$ and $H_V$, this ratio remains small prior to $\sim 40$ hr, when both helicity components are still weak. Beyond this stage, it increases steadily, reflecting the development of non-potential structures in the corona. Its temporal evolution closely resembles that of the free magnetic energy: both remain low initially and rise thereafter, although their absolute values differ.

At the time of observed eruption, the helicity ratio reaches 0.23, consistent with previous data-driven magnetofrictional studies reporting eruption thresholds of 0.17--0.40 for different ARs \citep{Pomoell2019, Price2019, Vemareddy2026}. However, the critical helicity ratio associated with eruption is not universal and depends on the magnetic configuration of the AR, its evolutionary history, and the method used to construct 3D magnetic fields. Previous investigations based on MHD simulations, magnetofrictional simulations, and NLFFF extrapolations have reported threshold values spanning $\sim0.1$--$\sim0.4$ \citep{Pariat2017, Zuccarello2018, Pomoell2019, Price2019, Thalmann2019, Thalmann2021, Gupta2021, Vemareddy2026}.

Although the helicity ratio reaches $\sim 0.23$ at the onset of the eruption, Figure~8 indicates that the FR remains relatively low in the corona, with an apex height of approximately 16~Mm. The helicity ratio subsequently increases to $\sim 0.29$ by 53~hr while the FR continues its gradual ascent. At $\sim 55$~hr, the FR apex enters the torus-unstable regime, where the decay index exceeds the critical threshold (described in the following section 3.4). As the FR rises further and reaches a height of $\sim 30$~Mm, the helicity ratio attains a value of $\sim 0.35$. These results suggest that the eruption is associated with the simultaneous increase in FR height and helicity ratio during the interval 48--60~hr, corresponding to a helicity-ratio range of approximately $0.23$--$0.35$.

It is important to note that the FR forms through the emergence of flux within pre-existing AR magnetic fields, and that a significant portion of the coronal structure is not part of the FR system (See Figure~\ref{fig:fieldlines1}). Including such structures in the helicity calculation can lead to variations in the inferred threshold values. For instance, in the AR 13500, \citet{Vemareddy2026} reported a helicity threshold of 0.29, consistent with FR heights in the corona and supporting the torus instability regime, in agreement with the threshold ratio proposed by \citet{Zuccarello2018}.

\subsection{Decay-index analysis and eruption mechanism}
We further inspected the eruption evolution in AIA 131~\AA, 171~\AA, and 304~\AA\ observations together with the STEREO-A and SolO perspectives. Throughout the rise phase, the erupting structure does not exhibit any signatures of writhing, helical deformation, or apex rotation that are typically associated with kink instability. Although localized regions of enhanced twist are present within the FR cross-section, the absence of clear kink-like morphological signatures suggests that kink instability is unlikely to be the dominant cause for triggering the eruption in this event.

Then, following previous observational and simulation studies \citep{DjPrice_2020, Mitra2022, Gupta2024}, we evaluated the decay-index profile deduced above the main PIL. The horizontal magnetic field component, $B_h=\sqrt{B_x^2+B_y^2}$, was averaged over localized regions surrounding the FR channel, and the decay index was computed as $n=-\frac{z}{{B}_{h}}\frac{\partial {B}_{h}}{\partial z}$. In the idealized model of \citet{Kliem2006}, a toroidal current ring embedded in an external magnetic field becomes torus-unstable when the decay index exceeds a critical value of $n_{crit}=1.5$. However, different values of the critical decay index have been reported across theoretical, numerical, and observational studies, reflecting the dependence of the instability threshold on the geometry and aspect ratio of the FR (e.g., \citealt{demoulin2010criteria, Gupta2024}). Notably, \citet{Zuccarello2015} suggested, based on numerical simulations, that the onset of torus instability occurs over a range of $n_{crit}\sim1.3-1.5$ rather than at a single unique threshold value, with the precise value depending on the specific magnetic configuration considered. In the present study, we adopt $n_{crit}=1.5$ as a reference threshold, while acknowledging that the onset of torus instability may occur at somewhat lower values of $n$ depending on the FR geometry \citep{Vasantharaju2019_decay}.

The resulting decay-index evolution is presented in Figure~\ref{decay_index} and compared with the temporal evolution of the modeled FR apex height. The critical height corresponding to the torus-instability threshold ( $n=1.5$) increases gradually from 22.94~Mm at 50~hr to 33.8~Mm at 70~hr, while the FR apex height rises from 16 Mm to 44.7 Mm during the same interval. At 50~hr, the FR remains below the critical torus height, indicating that the overlying coronal magnetic fields still provide sufficient confinement. However, by 55~hr, the FR apex height (26~Mm) exceeds the critical height (25~Mm), after which the structure continues to rise further into the torus-unstable regime. This behavior is consistent with the findings of \citet{Gupta2024}, who demonstrated, using a sample of ten large solar flares, that eruptive events are systematically associated with lower critical heights than confined ones, with eruptive flares in their sample all showing pre-flare averaged $h_\mathrm{crit}$ below $\sim$ 42~Mm. Moreover, \citet{Gupta2024} showed that once the FR altitude approaches the torus-unstable regime, the rapidly decreasing strapping field can no longer balance the hoop force of the current-carrying structure, facilitating the onset of eruption. Our results are consistent with this picture and with previous studies showing that eruptions occur when the restraining overlying magnetic fields decrease sufficiently rapidly with height \citep[e.g.][]{Zuccarello2015, Zuccarello2018, DjPrice_2020, Mitra2022, Gupta2024}, suggesting that torus instability is the most likely eruption mechanism in the present event. The onset of the torus-unstable regime at approximately 55 hr of evolution coincides with the helicity ratio reaching 0.32. Although the critical helicity ratio is not expected to be universal, this value provides an additional constraint on the eruptive threshold inferred for the present event.


\section{Summary and conclusion}

By using the photospheric magnetic fields and coronal EUV imaging observations, we studied the erupting scenario of the AR 12975. The AR had a simple bipolar configuration involving new bipolar flux regions emerging from March 27. These newly emerging flux regions developed through shear motions associated with the formation of the filament channel, as observed roughly one day prior to the eruption on March 28 at 12:00 UT. The emergence process injected significant magnetic helicity of $\approx 2.3 \times 10^{42}$~Mx$^{2}$ and magnetic energy of $\approx 2.5 \times 10^{32}$~erg into the corona by the eruption time at 12:00~UT on March 28, 2022 building a non-potential magnetic configuration. 

To model the observed magnetic evolution, we performed a time-dependent, data-driven MF simulation for three days starting from 12:00 UT on March 26. The initial potential field is driven by a photospheric electric field derived from a time sequence of vector magnetograms with an added non-inductive component \citep{Pomoell2019, Vemareddy_2024} parameterized to match the observed energy injection rate. The simulated structure is compared with the EUV observations at different stages, which delineates that the simulated magnetic evolution follows the coronal plasma tracers consistently.  In particular, the model reproduced the gradual formation of a twisted flux rope linked to the emerging flux regions over approximately 50~hours. The twisted structure is a sigmoidal FR that has remarkable morphological similarity observed in AIA 304~\AA \ observations from earth view and in EUI/171\AA\ images from the perspective view of SolO.  

\begin{figure*}[ht]
    \centering
    \includegraphics[width=0.9\textwidth]{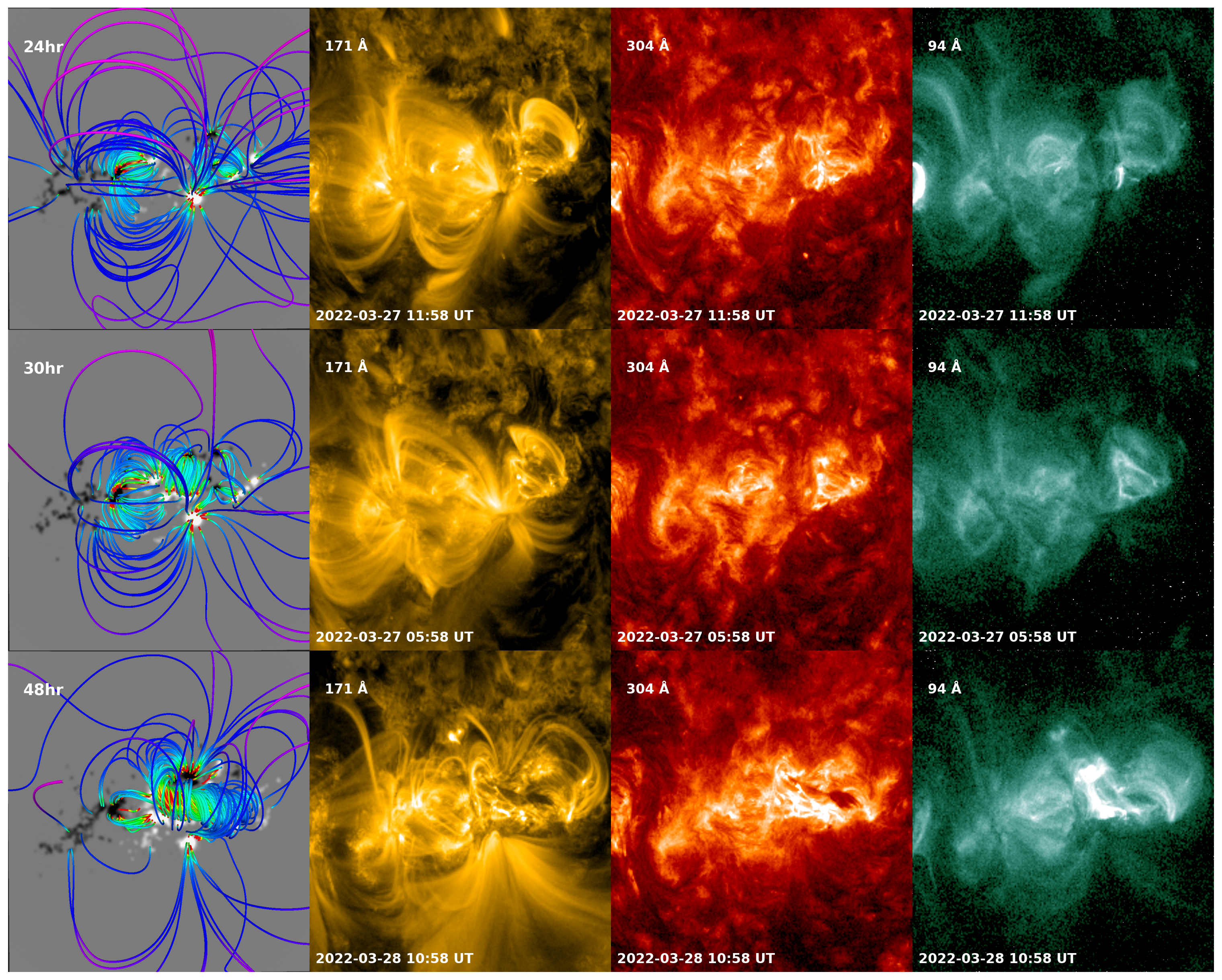}
    \caption{Comparison of the simulated structure with the observations. First column panels show top view of the simulated coronal magnetic fields at 24 hr, 30 hr, and 48 hr. Traced magnetic field lines are overplotted on a grayscale map of the vertical magnetic field component, $B_{z}$, illustrating the progressive development of a sheared and sigmoidal magnetic configuration. {\bf Second, third and fourth column panels} display the corresponding SDO/AIA observations in the 171~\AA, 304~\AA, and 94~\AA\ channels respectively. }
    \label{fig:fieldlines1}
\end{figure*}

\begin{figure*}[ht]
    \centering
    \includegraphics[width=0.99\textwidth]{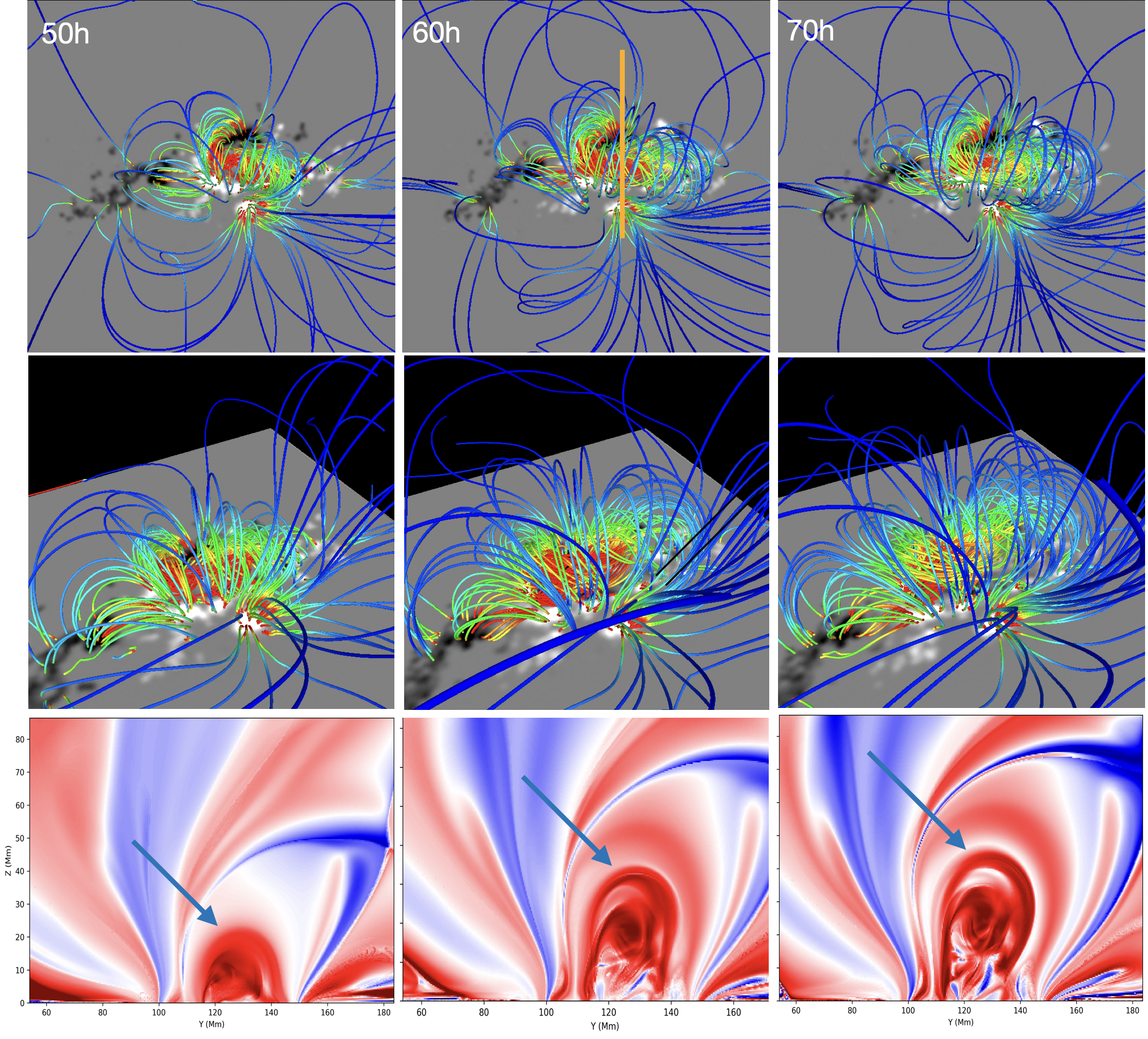}
    \caption{Rendered magnetic structure at 50hr, 60hr, 70hr in the top view in the first row, and perspective view in the second row panels. The structure mimics the sigmoid which becomes prominent as it builds up.  Third row panels display the field line twist derived in a plane placed across the FR (vertical yellow line) which discern the upward rise motion of the sigmoidal FR. Red (blue) color refers to a negative (positive) twist scaled to 2 turns. The arrow points to the FR cross-section. }
    \label{fig:fieldlines2}
\end{figure*}

\begin{figure*}[!htb]
    \centering
    \includegraphics[width=0.8\textwidth]{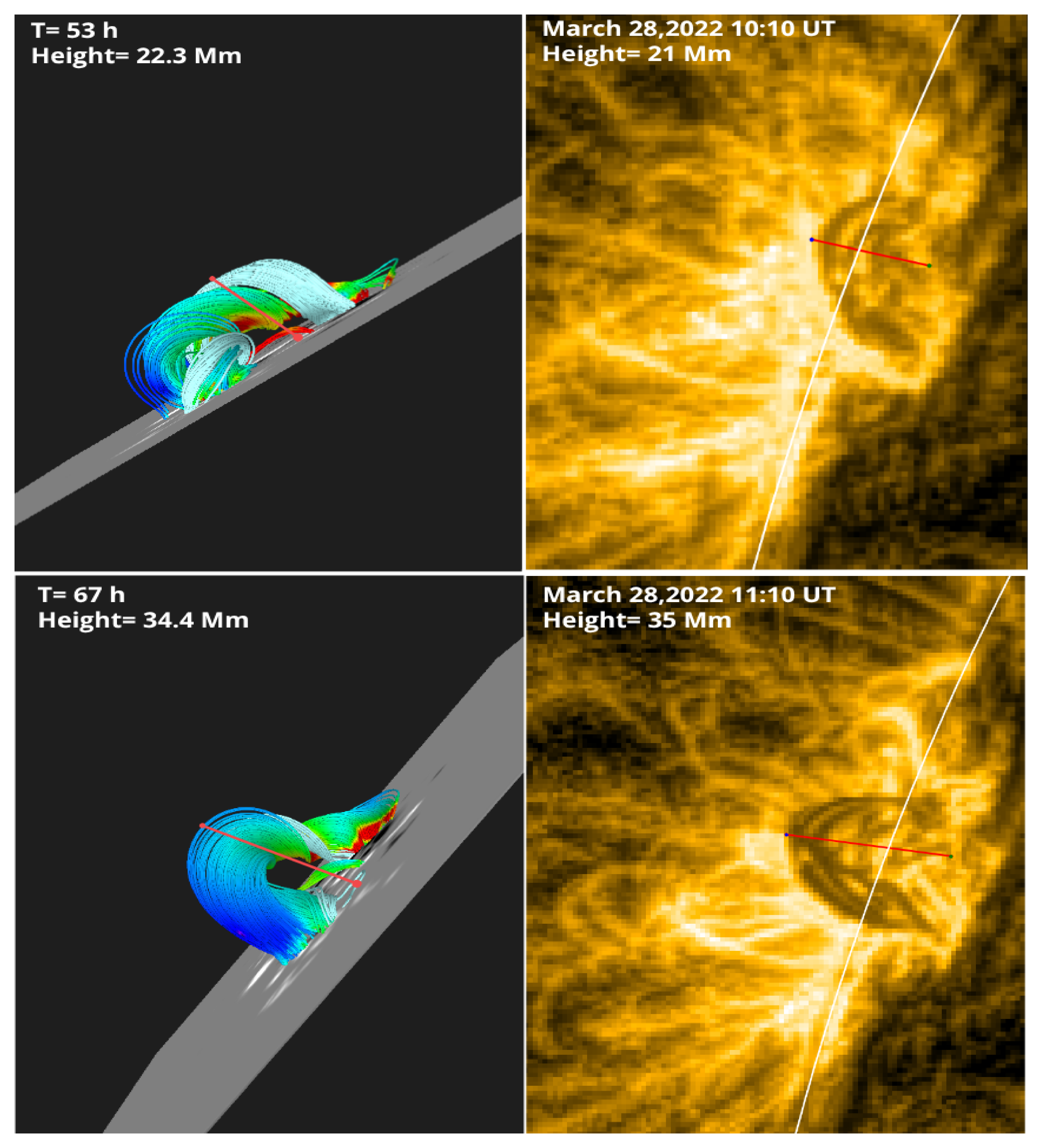}
    \caption{Comparison of the modeled FR with SoLO observations. \textbf{Left panels}: The magnetic field lines from the simulation showing the FR structure at two representative times (top: $t = 53$~hr; bottom: $t = 67$~hr). The red line indicates the direction along which the apex height of the FR is measured. The corresponding heights are $\sim 22.3$~Mm and $\sim 34.4$~Mm, respectively. \textbf{Right panels}: Limb-view EUV images from SolO at 10:10~UT and 11:10~UT on March 28, 2022 showing the erupting filament. The projected filament height, measured along the same direction (red line), increases from $\sim 21$~Mm to $\sim 35$~Mm. }
    \label{fig:filament_height}
\end{figure*}

\begin{figure*}[!htb]
    \centering
    \includegraphics[width=0.7\textwidth]{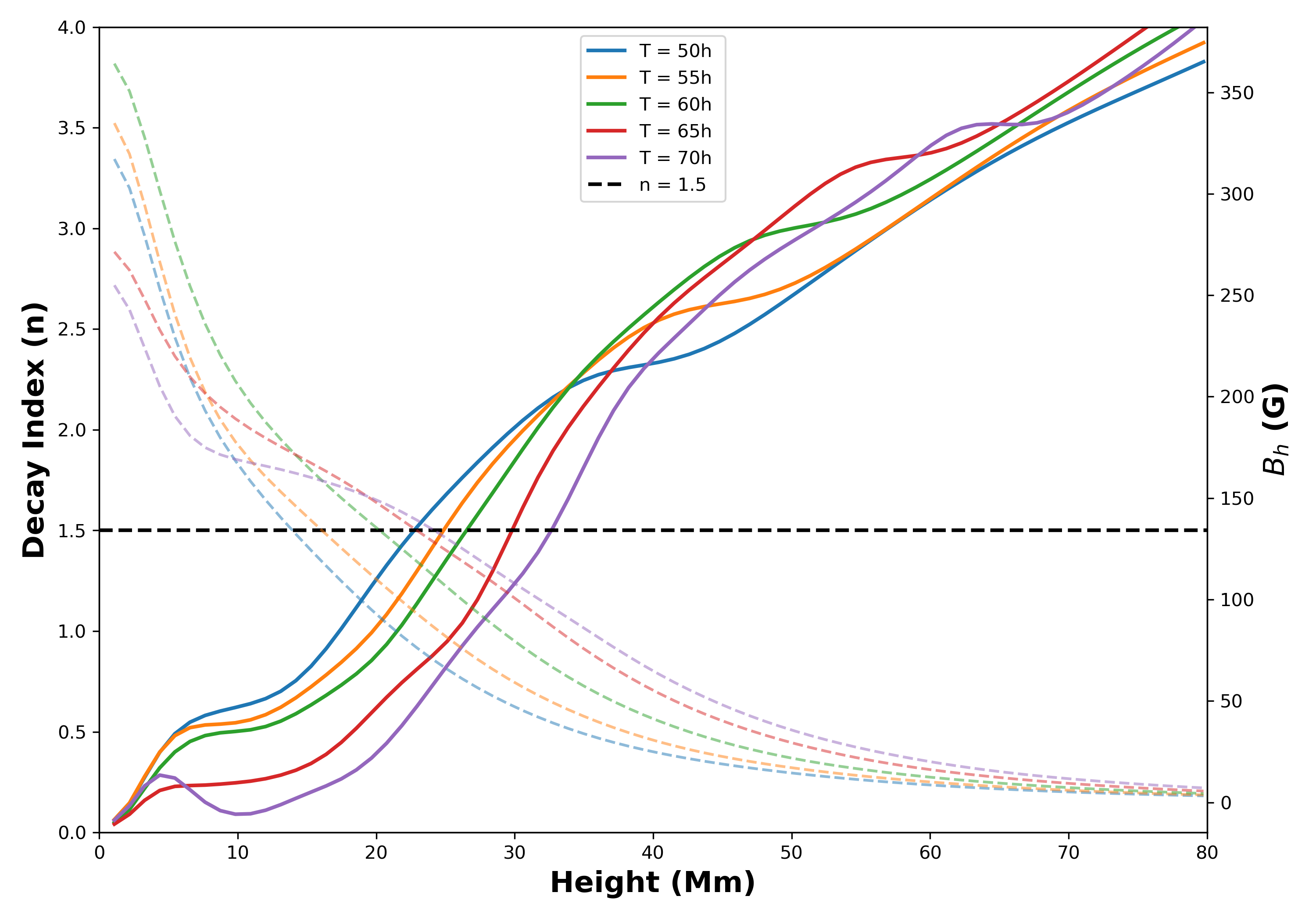}
     \caption{Decay-index profiles above the polarity inversion line (PIL) at different simulation times ($T=50$, 55, 60, 65, and 70 hr). The solid curves represent the decay index n, while the dashed curves show the corresponding horizontal magnetic field strength $B_{h}$ as a function of height. The horizontal dashed black line marks the torus-instability threshold ($n=1.5$). The gradual rise of the flux rope above the critical height indicates the onset of torus instability during the eruption evolution.}
    \label{decay_index}
\end{figure*}

\begin{figure*}[!ht]
    \centering
    \includegraphics[width=0.7\textwidth]{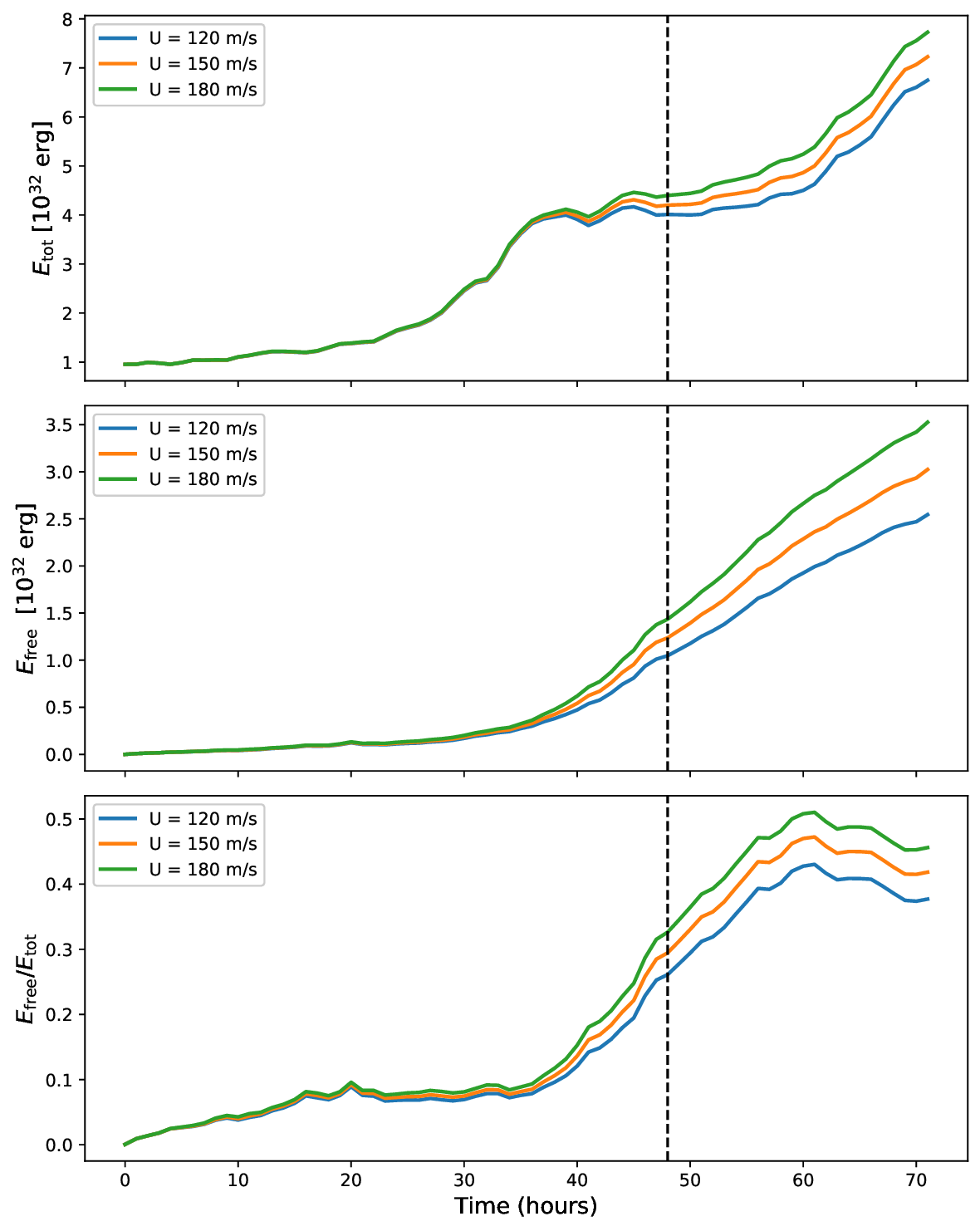}
     \caption{Time evolution of magnetic quantities for the three flux-emergence driving speeds: 120 m/s (blue), 150 m/s (orange), and 180 m/s (green).  \textbf{Top panel}: Total magnetic energy $E(t)$ in units of $10^{32}$ erg.
    \textbf{Middle panel}: Free magnetic energy $E_{\rm free} = E - E_{\rm p}$, where $E_{\rm p}$ is the potential-field energy.\\  \textbf{Bottom panel}: Fraction of free magnetic, $E_{\rm free}/E_{\rm tot}$. Vertical dashed lines denote the onset time of the eruption. }
    \label{fig:energy_metric}
\end{figure*}

\begin{figure*}[!ht]
    \centering
    \includegraphics[width=0.7\textwidth]{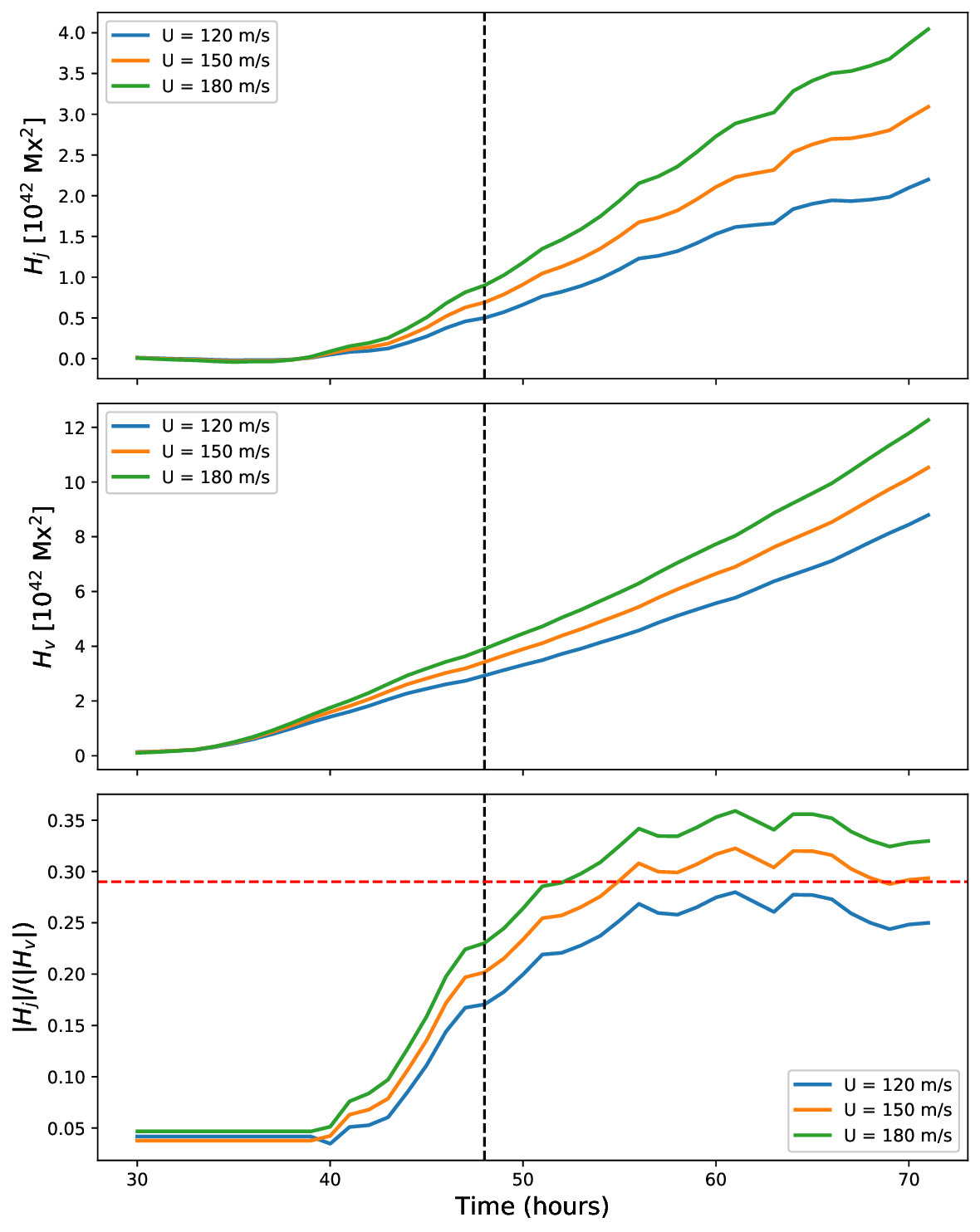}
     \caption{Time evolution of magnetic quantities for the three flux-emergence driving speeds: 120 m/s (blue), 150 m/s (orange), and 180 m/s (green). \textbf{Top panel}: Current-carrying helicity $H_{j}(t)$ (in units of $10^{42}$ Mx$^2$). \textbf{Middle panel}: Total relative magnetic helicity $H_{V}$ in the coronal volume (in units of $10^{42}$ Mx$^2$).\\
    \textbf{Bottom panel}: Ratio $|H_{\rm j}|/|H_{V}|$ (Bottom panel) indicating the fractional contribution of the non-potential helicity to the total helicity. Vertical dashed lines denote the onset time of the eruption.}
    \label{fig:helicity_ratio}
\end{figure*}

The simulated time evolution of magnetic energy and helicity in the computational volume corresponds with the observed injection of energy and helicity. Both quantities exhibit a steady increase, reflecting continuous injection from the photosphere. The free magnetic energy reaches approximately $1.5 \times 10^{32}$~erg, corresponding to about $35\%$ of the total magnetic energy prior to eruption, consistent with values reported in previous TMF studies \citep{Pomoell2019, Price2019, DjPrice_2020, Vemareddy_2024}. Keeping in view of the FR heights (see Figure 8), we suggest that an eruption is likely within the time window of 48–60 hr, corresponding to a ratio $\left|H_j\right|/\left|H_V\right|$ range of 0.23–0.35. However, based on the combined analysis of the eruption mechanism, the filament became susceptible to torus instability after approximately 55 hours of evolution, at which point the helicity ratio reached 0.32. This value lies well within the previously reported range of eruption thresholds 0.17–0.40 observed across different ARs \citep{Pomoell2019, Price2019, Vemareddy2026}. It is important to note that the FR forms within pre-existing magnetic fields in this AR, and that a significant portion of the coronal structure is not part of the FR system (see Figure~\ref{fig:fieldlines1}). As a result, the computed helicity ratio yields values that significantly differ from the threshold proposed by \citet{Zuccarello2018}.


Another difficulty with our data-driven model is the amount of energy injection that is parameterized according to the observed one inferred from photospheric velocity fields. The parameter that determines the non-inductive contribution has to be as high as twice the observed one in order to form comparable twisted structures \citep{Vemareddy_2024}. This is probably due to weak horizontal fields that are less sensitive in the regions linked to the FR. This higher value of the non-inductive parameter is not required in some ARs. The AR 13500 is an example where the observed injection of energy simulated the magnetic evolution, capturing the FR (filament/sigmoid) formation, and the helicity threshold of 0.29 meets the torus height \citep{Vemareddy2026}, as conjectured by \citet{Zuccarello2018}. 

More generally, eruptions are observed in diverse ARs where the erupting coronal structures often coexist with adjacent flux systems that are not directly involved in the eruption. The present AR represents such a case, where the FR forms within emerging weak-field regions. In particular, the horizontal magnetic fields do not adequately capture the electric currents required for accurate extrapolation, which are essential for reproducing the observed filament. By allowing enhanced energy injection, our data-driven MF approach successfully reproduces filament formation with a high degree of morphological fidelity. 

\begin{acknowledgments}
We are grateful to the anonymous reviewer for the constructive comments and suggestions, which significantly improved the scientific quality and clarity of this work. We acknowledge the use of data from the \textit{Solar Dynamics Observatory} (SDO), including the \textit{Helioseismic and Magnetic Imager} (HMI) and the \textit{Atmospheric Imaging Assembly} (AIA), as well as complementary observations from \textit{SOHO/LASCO}, \textit{Solar Orbiter}, and \textit{STEREO/SECCHI}. HMI vector magnetogram data were obtained from the \textit{Joint Science Operations Center} (JSOC), and GOES soft X-ray flux data were used for flare identification. The numerical simulations were performed using the \textit{PENCIL Code} in a magnetofrictional setup, and magnetic field visualizations were produced with the \textit{VAPOR} software (\url{https://www.ucar.edu/vapor}). This work was supported by the \textit{Department of Space, and Department of Science and Technology, Government of India}.

\end{acknowledgments}


\section{Data Availability}

All data used in this study are publicly available. SDO/HMI and SDO/AIA data were obtained from the \textit{Joint Science Operations Center} (JSOC; \url{http://jsoc.stanford.edu}). GOES soft X-ray flux data were accessed via the NOAA data archive. Solar Orbiter data were obtained from the ESA \textit{Solar Orbiter Archive} (SOAR; \url{https://soar.esac.esa.int/soar/}). STEREO/SECCHI data were obtained from the STEREO Science Center (\url{https://stereo-ssc.nascom.nasa.gov/data.shtml}). Complementary coronagraph observations from \textit{SOHO/LASCO} are available through the SOHO data archive.

\section{ORCID iDS}
Dinesh Mishra \url{https://orcid.org/0009-0000-3793-0779}
P. Vemareddy \url{https://orcid.org/0000-0003-4433-8823}
Brajesh Kumar \url{https://orcid.org/0000-0002-7054-5669}
\bibliography{references}{}
\bibliographystyle{aasjournalv7}



\end{document}